\documentclass[twocolumn,aps,prl,superscriptaddress]{revtex4-1}

\usepackage{graphicx}
\usepackage{bm}
\usepackage{multirow}

\begin{document}

\title{Double ellipsoidal Fermi surface model of the normal state of ferromagnetic  superconductors}


\author{Christopher L\"{o}rscher }

\affiliation{Department of Physics, University of Central Florida, Orlando, FL 32816-2385 USA}
\author{Jingchuan Zhang}
\affiliation{Department of Physics, University of Central Florida, Orlando, FL 32816-2385 USA}
\affiliation{Department of Physics, University of Science and Technology Beijing, Beijing 100083, China}
\author{Qiang Gu}
\affiliation{Department of Physics, University of Science and Technology Beijing, Beijing 100083, China}
\author{Richard A. Klemm}
\affiliation{Department of Physics, University of Central Florida, Orlando, FL 32816-2385 USA}


\date{\today}

\begin{abstract}
We  model the normal state of ferromagnetic superconductors with two general ellipsoidal Fermi surfaces (FSs), one for each spin projection $\sigma=\{\uparrow,\downarrow\}$, each with its ferromagnetically split chemical potential $\mu_{\sigma}$ and its three distinct single particle effective masses, $\{m_{i\sigma}\}$, the geometric mean of which is $m_{\sigma}$. We study this  model in the presence of an arbitrarily oriented magnetic induction, ${\bm B}=\mu_{0}{\bm H}+{\bm M_{0}}$, where ${\bm M_{0}}$ includes the Ising-like spontaneous ferromagnetic order, which for URhGe is in the $c$-axis direction above the superconducting transition temperature $T_c$.  In analogy to the Sommerfeld low-temperature $T$ expansion with $B=0$, we assume the low-$T$ total particle density  $\Sigma_{\sigma} n_{\sigma}({\bm B})$ to be independent of ${\bm B}$,  and obtain a self-consistent asymptotic expansion for $\sum_{\sigma}\Pi^{3/2}_{\sigma}({\bm B})$ in even powers of ${\bm B}$, where  $\Pi_{\sigma}({\bm B})=m_{\sigma}({\bm B})\mu_{\sigma}({\bm B})$.  We assume that the $\mu_{\sigma}({\bm B})$ are linear in ${\bm B}$ for both spins due to the Zeeman interaction and that the remaining even ${\bm B}$ dependence in the $\Pi_{\sigma}({\bm B})$ arises only from  $m_{\downarrow}({\bm B})$.  An analogous procedure leads to an asymptotic expansion in even powers of ${\bm B}$ for the  linear $T$-coefficient, $\gamma({\bm B})$, of the low-$T$ specific heat $C_{V}({\bm B})$. Our expression for $\gamma({\bm B})$ leads to  good  fits to the $\gamma({\bm H})$ data of Aoki and Flouquet [J. Phys. Soc. Jpn. \textbf{81}, 011003 (2012)] obtained for the ferromagnetic superconductor URhGe in the ferromagnetic, non-superconducting phase, with the applied magnetic field ${\bm H}$ along each of the three crystallographic directions. We discuss this model in terms of the reentrant superconducting properties of URhGe and UCoGe. This model can be generalized to an arbitrary number of ellipsoidal FSs.

\end{abstract}


\maketitle


Recent discoveries of heavy fermion superconducting materials such as UGe$_{2}$ \cite{Saxena}, UCoGe \cite{Huy,Hattori}, and URhGe \cite{Aoki1,Hardy, Levy1, Yelland, Levy2, Aoki2} in which there is simultaneous ferromagnetic and superconducting order, have sparked renewed interest in the field of $p$-wave superconductivity.  For such superconductors, the symmetry of the spin component of the wave function is odd ({\it i.e.} $l=1,3,...$), with $p$-wave symmetry ($l=1$) being the simplest example of such a case. In these novel superconductors, the Cooper spin pairs form triplet states, as opposed to singlet states that their $s$-wave counterparts form. The parallel-spin triplet states are much more resilient to the externally applied magnetic field ${\bm H}$, which is evident from (1) their unusually high zero-temperature upper critical inductions, $B_{c2}(0)$, which in some cases exceeds the Pauli limit $B_P$ by a factor of twenty in at least one crystallographic direction, where $B_{P} \sim 1.85T_{c}$ T/K, where $T_{c}$ is the superconducting transition temperature in K, and (2) by the temperature $T$ independence of the  Knight shift for applied fields ${\bm H}$ normal to the direction of the ferromagnetism\cite{Hattori}, so that these experiments appear to be consistent with one another. In contrast, the Knight shift and $B_{c2,||}(0)$ for fields parallel to the layers of Sr$_2$RuO$_4$, are inconsistent with one another\cite{book}. In the magnificent case of URhGe, there has been strong evidence of an anomalous high-field reentrant  superconducting phase measured in clean samples with residual resistance ratio RRR = 50 \cite{Levy2}, where the superconductivity was found to disappear at a relatively low field strength \cite{Hardy}, but then reappears when the strength of the external field exceeds 8 T\cite{Levy1}. The robustness of the superconductivity in high fields is a signature of  parallel-spin pairing, and cannot be easily  explained using conventional BCS $s$-wave pairing.  Although Knight shift measurements have  not yet been performed in either the high or low-field superconducting states of URhGe, its similarity to its sibling ferromagnetic superconductor UCoGe strongly suggests that it is also a parallel-spin superconductor, most likely with a $p$-wave polar state fixed to the crystallographic $a$-axis direction\cite{Hardy,SK2}.

In addition to the mysterious reentrant superconductivity observed in clean samples of URhGe, Shubnikov de Haas (SdH) measurements were also performed on even cleaner samples of URhGe with an RRR=130, from which Yelland {\it et al.} observed a sudden disappearance of SdH oscillations in the field-dependent resistance, $R({\bm B})$ for $\mu_{0}\bm H \geq 15.5$ T \cite{Yelland}. Those authors claimed the disappearance of the SdH oscillations was due to a topological Lifshitz Fermi surface (FS)  transition, where the field-dependent cross-sectional area of the FS, $A({\bm B})$, suddenly vanishes, quenching the SdH oscillations. They attributed this effect partially to a decrease in the effective cyclotron mass $m^{*}$, but primarily to a decrease in the field-dependent Fermi velocity, ${\bm v}_F({\bm B})=\hbar k_F/m^{*}$ with a smooth drop to zero at around 15 T in the wave vector $k_F({\bm B})$. Yelland {\it et al.} also claimed that a strong ${\bm H}||\hat{\bm b}$  increases the pairing interaction strength $V_0$ and decreases the  effective ${\bm v}_F({\bm B})$  of the heavy-electron ellipsoidal FS responsible for the pairing \cite{Yelland,Davis,Shick,Miiller}.  Yelland {\it et al.} further attributed the dramatic re-entrance into the superconducting state of URhGe at high magnetic fields in the $b$-direction as a direct consequence of this.  However, a similar effect in UCoGe was claimed to be due to anomalies in the effective mass $m({\bm B})$\cite{Aoki2}.  Thermopower measurements at large fields provided strong evidence for a change in the FS in UCoGe\cite{Malone1}, but no such change at the reentrant field in URhGe\cite{Aoki_Flouquet,Malone2}.  As noted in the most recent review article on the subject, it is presently unclear as to whether the FS changes dramatically with ${\bm H}$, as in a vanishing of the average Fermi wavevector $k_F$, or whether the effective mass is strongly enhanced with  ${\bm H}$\cite{Aoki3}.  Anomalously anisotropic magnetization ${\bm M}({\bm H})$ measurements of the $T$ derivative $\gamma({\bm H})$ of the specific heat in URhGe and of the coefficient $A$ in the low-$T$ resistivity $\rho(T)=\rho_0+AT^2$ in UCoGe were claimed to support the latter interpretation\cite{Aoki2,Aoki3}, the latter using the Kadowaki-Woods relation\cite{Kadowaki}.  But all of these works assumed a spherical FS, which for orthorhombic UCoGe and URhGe is certainly not the case\cite{Aoki_Flouquet}.  Thus, if the enhancement of the ``effective mass'' with ${\bm H}$ actually occurs, one should try to determine which of the effective masses on which of the relevant FSs shows the strong enhancement.  In the following, we show  that strong changes with applied field occurring on only  one of the three single particle effective masses on one of the ellipsoidal FSs in our double ellipsoidal FS model can explain the specific heat data on URhGe.

To help resolve this controversy, low-$T$ ${\bm M}({\bm H})$ measurements on URhGe yielded the slope  $\gamma({\bm H})$ with $T$ of the specific heat for ${\bm H}||\hat{\bm a},\hat{\bm b},\hat{\bm c}$ from the Maxwell relation \cite{Aoki2}.  They found that $\gamma({\bm H})$ remains relatively flat for ${\bm H}||\hat{\bm a}$ with a slight hint of upward curvature, decreases approximately linearly for ${\bm H}||\hat{\bm c}$ with slight upward curvature, but increases with ${\bm H}||\hat{\bm b}$ for $H<<H_R$, the ``reentrant field'',  up to a sharp maximum at $\mu_{0}{\bm H}\cdot\hat{\bm b}\sim~12~$T$~\sim~\mu_{0}H_R$, then decreases to  a value higher than that at ${\bm H=0}$.

In this paper we analytically calculate $\gamma({\bm B})=\partial(S/V)/\partial T$ for an electron gas with two ferromagnetically split ellipsoidal Fermi surfaces, one for each spin projection, $\sigma=\{\uparrow,\downarrow\}$, with three distinct single particle effective masses, $\{m_{i\sigma}\}$, describing each FS.  We study this double ellipsoidal FS model in the presence of an arbitrarily oriented magnetic induction, ${\bm B}=\mu_{0}{\bm H}+{\bm M_{0}}$, and qualitatively compare our results to the experimental curves of $\gamma({\bm B})$ for all three crystallographic directions measured for URhGe.

\begin{figure}
\center{\includegraphics[width=0.3\textwidth]{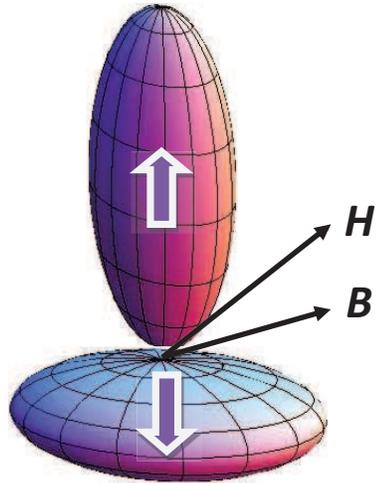}
\caption{(color online) Plot of two ellipsoidal Fermi surfaces aligned along the crystal axes corresponding to the up and down electron spin states.  The applied magnetic field ${\bm H}$ and the magnetic induction ${\bm B}$ are indicated by the arrows.}}
\end{figure}

\begin{figure}
\hskip20pt\center{\includegraphics[width=0.3\textwidth]{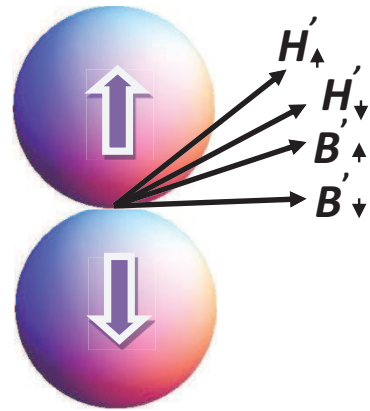}
\caption{(color online) Plot of the ellipsoidal Fermi surfaces aligned along the crystal axes corresponding to the up and down electron spin states after transforming them to spherical shapes.  The transformed applied magnetic fields ${\bm H}'_{\uparrow}$ and  ${\bm H}'_{\downarrow}$ and the transformed magnetic inductions ${\bm B}'_{\uparrow}$  and ${\bm B}'_{\downarrow}$  are in general different on each Fermi surface.}}
\end{figure}

We begin our calculation with the Hamiltonian of the system
\begin{equation}\label{eq1}
H=\sum_{\sigma=\uparrow, \downarrow}\epsilon_{\sigma}({\bm{k}}-e{\bm{A}})-\mu_{\sigma}({\bm B}),
\end{equation}

where
\begin{equation}\label{eq2}
\mu_{\sigma}({\bm B})=\mu(0)+\sigma I/2-g\mu_{B}{\bm\sigma}\cdot {\bm B}/2,
\end{equation}

and
\begin{equation}\label{eq3}
\epsilon_{\sigma}({\bm{k}})=\sum_{i=1}^{3}k_{i}^{2}/[2m_{i\sigma}({\bm B})],
\end{equation}
$\mu_{B}$ is the Bohr magneton, $g \sim 2$, $g\mu_{B}/2=\gamma$ is the gyromagnetic ratio for the electron, $\sigma=\{\uparrow,\downarrow\}$, the $\mu_{\sigma}(\boldsymbol{B})$ are the chemical potentials  on each FS at $T=0$ in the magnetic induction, $\boldsymbol{B}$, $\mu(0)$ is the non-magnetic chemical potential at $T=0$ and ${\bm B=0}$, ${\bm A}$ is the magnetic vector potential, $e$ is the magnitude of the charge of an electron, $I$ is the Stoner coupling energy, ${\bm{B=\nabla\times A}}$,  the $m_{i\sigma}({\bm B})$ are the induction dependent single particle effective masses on each FS, and we set $\hbar=1$.

Here, we included the effects of two distinct ellipsoidal Fermi surfaces, which are split by the ferromagnetism,  one for each spin projection, by including three single particle effective masses for each spin projection, $m_{i\sigma}({\bm B})$. Figure 1 depicts the two distinct FSs which are aligned along the three crystal axis directions, with the externally applied magnetic field, $\bm H$, applied in some general direction, and the magnetic induction, $\bm B$, which includes the spontaneous magnetization, $\bm M_{0}$. Our model can naturally be extended to include any arbitrary number of FSs. We perform the first of the Klemm-Clem (KC)-transformations \cite{KC,book,Lorscher} on the two distinct spin-split FSs, $x_{i}=\overline{m}^{-1/2}_{i\sigma}x_{i\sigma}^{\prime}$, {\it etc.}, to map both ellipsoidal FSs onto spherical ones, where $\overline{m}_{i\sigma}=\frac{m_{i\sigma}({\bm B})}{m_{\sigma}({\bm B})}$, and $m_{\sigma}({\bm B})=[m_{1\sigma}({\bm B})m_{2\sigma}({\bm B})m_{3\sigma}({\bm B})]^{1/3}$ is the geometric mean effective mass on the $\sigma$ FS.  Figure 2 qualitatively depicts the two spherical FSs after the scale transformations are performed. Note that the transformations performed for the $\sigma$ FS depend on the single particle effective masses $m_{i\sigma}$ relevant for that FS, and that each transformation changes the effective directions of ${\bm H}$ and ${\bm B}$ to ${\bm H}^{\prime}_{\sigma}$ and ${\bm B}^{\prime}_{\sigma}$ for the $\sigma$ FS, which are different on the $\sigma = \uparrow,\downarrow$ FSs.  In Fig. 2, the differently transformed fields are indicated by the arrow subscripts. We then rotate these transformed fields to the crystal $z^{\prime}=z$-axis direction, and finally apply isotropic scale transformations involving the anisotropy parameter $\alpha_{\sigma}\left(\theta,\phi\right)$ and obtain expressions for the transformed angles $\mathrm{cos}\theta_{\sigma}^{\prime}=\sqrt{\overline{m}_{3\sigma}}\frac{\mathrm{cos}\theta}{\alpha_{\sigma}\left(\theta,\phi\right)}$, $etc.$, where
\begin{eqnarray}\label{eq15}
\alpha_{\sigma}\left(\theta,\phi\right)&=&[\overline{m}_{1\sigma}\mathrm{sin}^{2}\theta\mathrm{cos}^{2}\phi+\overline{m}_{2\sigma}\mathrm{sin}^{2}\theta\mathrm{sin}^{2}\phi\nonumber\\
&&+\overline{m}_{3\sigma}\mathrm{cos}^{2}\theta]^{1/2}.
\end{eqnarray}

We also obtain $\epsilon_{\sigma}(k_{\sigma,\parallel},n+1/2)=\frac{k_{\sigma,\parallel}^{2}}{2m_{\sigma,\parallel}}+\omega_{\sigma,\perp}(n+1/2)$, $\omega_{\sigma, \perp}=\frac{eB}{m_{\sigma,\perp}({\bm B})}$, $m_{\sigma, \perp}({\bm B})=\frac{m_{\sigma}({\bm B})}{\alpha_{\sigma}\left(\theta,\phi\right)}$, and $m_{\sigma,\parallel}=m_{\sigma}\alpha_{\sigma}^{2}(\theta,\phi)$ for the effective masses parallel and perpendicular to the transformed magnetic inductions on each of the two transformed FSs.  Although obtaining the effective mass parallel to the field with the KC-transformations is non-trivial, we have shown that our results are consistent for any choice of $\boldsymbol{A}$ in all three crystallographic planes. More details are presented in the appendix.

We begin with the expression for the total particle density for both spin-split Fermi surfaces in the presence of $\bm B$, $\Sigma_{\sigma} n_{\sigma}({\bm B})$, which we assume to be independent of ${\bm B}$,  in analogy with the Sommerfeld low temperature $T$ expansion with $B=0$. We obtain an asymptotic series expression involving the products $\Pi_{\sigma}(\bm B)=m_{\sigma}(\bm B)\mu_{\sigma}(\bm B)$ of the induction dependent chemical potential of each FS, $\mu_{\sigma}(\bm B)$, and the induction dependent geometric mean effective mass of each FS, $m_{\sigma}({\bm B})$,
\begin{eqnarray}\label{eq4}
&&\sum_{\sigma}\Pi_{\sigma}^{3/2}\left(0\right)=\sum_{\sigma}\Pi_{\sigma}^{3/2}\left(\bm B\right)\nonumber \\
&&\times\left(1+\sum_{n=0}^{\infty}a_{n}\left(\frac{eB\alpha_{\sigma}\left(\theta,\phi\right)}{\Pi_{\sigma}\left(\bm B\right)}\right)^{2n+2}\right),
\end{eqnarray}
where
\begin{equation}\label{an}
a_{n}=3(-1)^{n+1}(2-2^{-2n})\zeta(2n+2)\frac{(4n-1)!!}{(4\pi)^{2n+2}}.
\end{equation}
We then calculate the linear-$T$ coefficient of the induction dependent specific heat, $\gamma({\bm B})$. We define the entropy of the system in the usual way
\begin{equation}\label{eq5}
S=-k_{B}\sum_{\mathbf{k\sigma}}[n_{F\sigma}\mathrm{ln}n_{F\sigma}+\left(1-n_{F\sigma}\right)\mathrm{ln}\left(1-n_{F\sigma}\right)],
\end{equation}
where
\begin{equation}\label{eq6}
n_{F\sigma}=\frac{1}{e^{\beta\left[\epsilon_{\sigma}({\bm{k}}-e{\bm{A}})-\mu_{_{\sigma}}({\bm B})\right]}+1},
\end{equation}
$k_{B}$ is  Boltzmann's constant, and $\beta=\frac{1}{k_{B}T}$. The details of this calculation are presented in the appendix. Making use of the thermodynamic relation $C_{V}=-\beta\frac{\partial\left(S/V\right)}{\partial\beta}$, we obtain the $B=0$ linear $T$ coefficient of the specific heat,
\begin{equation}\label{eq7}
\gamma\left(0\right)=\frac{k_{B}^{2}}{3\sqrt{2}}\sum_{\sigma}m_{\sigma}\left(0\right)\Pi_{\sigma}^{1/2}\left(0\right),
\end{equation}
which agrees with textbook formulas.

For finite $\bm B$, the linear $T$-coefficient of the specific heat can be expanded in an asymptotic series
\begin{eqnarray}\label{eq8}
\gamma\left({\bm B}\right)&=&\frac{k^{2}_{B}}{3\sqrt{2}}\sum_{\sigma}m_{\sigma}\left(\bm B\right)\Pi_{\sigma}^{1/2}\left(\bm B\right)\nonumber\\
&&\times\left(1+\sum_{n=0}^{\infty}\lambda_{n}\left(\frac{eB\alpha_{\sigma}\left(\theta,\phi\right)}{\Pi_{\sigma}\left(\bm B\right)}\right)^{2n+2}\right),
\end{eqnarray}
where $\lambda_{n}=\frac{1}{3}\left(4n+1\right)a_{n}$.

To fit the specific heat data of Aoki $et$ $al.$ \cite{Aoki_Flouquet}, we first used a least-squares fit for the least anomalous $a$ and $b$-axis directions of ${\bm H}$ to obtain the explicit field dependence of $\gamma$. Since SdH oscillations were found to vanish for URhGe with increasing ${\bm H}||\hat{\bm b}$, we assume that either the FS  warps upon application of a strong magnetic field ${\bm H}$, or that the increased ${\bm H}$ moves the plane of electronic orbits away from  an optimal cross-sectional area $A$ of the FS, for which $\frac{\partial A}{\partial k_{\sigma,\parallel}}=0$, thus quenching the SdH oscillations. In this latter scenario, we assume that this change in the observed portion of the FS with applied field will introduce a field dependence to the single particle effective mass parallel to the $b$-axis direction on the dominant FS. We assume that the single particle effective mass parallel to the $b$-axis direction on the $\downarrow$ FS, which dominates the ${\bm H}\cdot\hat{\bm b}$ dependence of $\gamma$,  to have a Lorentzian field dependence of the form

\begin{equation}\label{eq9_new}
m_{b,\downarrow}(\bm B)=\sum_{\alpha=\pm1}\frac{m_{b,\downarrow}(0)\left(H_{0}^{2}+(\Delta H)^{2}\right)}{(H+\alpha H_{0})^{2}+(\Delta H)^{2}}+m_{b,\downarrow,\infty},
\end{equation}
where the fitting parameter $\mu_{0}H = 12.0~$T is the magnetic field strength at the onset of the metamagnetic transition signaled by the peak in the $\frac{\partial M}{\partial H}$ data and the $\gamma_{\bm H \parallel b}$ data.  The width $\mu_{0}\Delta H = 0.97~$T is proportional to the full width at half maximum of the peak, $m_{b, \downarrow}(0)$ is the effective mass along the $b$-axis direction on the $\downarrow$ FS in zero field, and $m_{b,\downarrow,\infty}\ll m_{b,\downarrow}(0)$ is the negligible additional contribution to the effective mass  along the $b$-axis direction on the $\downarrow$ FS at high fields well above the metamagnetic transition.  All other single particle effective masses on both FSs are assumed to be independent of field, and all other field dependencies arise from the field-dependent chemical potential of the dominant FS, $\mu_\downarrow\approx\mu_{\downarrow}(0)+\delta\mu_{\downarrow}(\mu_{0}H)$, where $\mu_{\downarrow}(0)$ and $\delta\mu_{\downarrow}$ are fitting parameters.

We substitute Eq. (11) into Eq. (10) with $\mu_\downarrow=\mu_{\downarrow}(0)+\delta\mu_{\downarrow}(\mu_{0}H)$ and include only the dominant $\downarrow$ FS, and expand to leading order, obtaining (in J/mol K$^2$),
\begin{eqnarray}
\gamma_{b}(H)&=&\frac{k_{B}^2}{2\hbar^{3}\sqrt{3}}\lambda\left[\mu_{\downarrow}(0)+\delta\mu_{\downarrow}(\mu_{0}H)\right]^{1/2}\\ \nonumber
&&\times \left(\sum_{\alpha=\pm1}\frac{m^{3}_{\downarrow}(0)\left(H_{0}^{2}+(\Delta H)^{2}\right)}{(H+\alpha H_{0})^{2}+(\Delta H)^{2}}\right)^{1/2}+\gamma_{b,\infty}
\end{eqnarray}
where $\lambda=3.4\times10^{-5}$ m\textsuperscript{3}/mol is the conversion factor we calculated based on 4 U atoms per unit cell \cite{Aoki_Flouquet}, the dimensions of an orthorhombic unit cell\cite{Tran}, the number of carriers per U atoms\cite{Yelland}, and Avogadro's number. The quantity $m_{\downarrow}(0)$ is the geometric mean mass for the $\downarrow$ FS in zero field.  In our fits to the $\gamma_b({\bm H})$, we again obtain $\mu_{0}H=12.0$ T and  $\mu_{0}\Delta H=0.97$ T, but we further obtain $m_{\downarrow}(0)\sim 180~m_{e}$, $\gamma_{b,\infty}=0.15~\textup{J/(mol K\textsuperscript{2})}$, $\mu_{\downarrow}(0) = 20~\textnormal{meV}$, and $\delta\mu_{\downarrow}=3.1~\textnormal{meV/T}$, where $m_e$ is the mass of an electron in vacuum. We note that the positive value for $\delta\mu_{\downarrow}$ we obtained supports our hypothesis that the $\downarrow$ FS is the dominant FS contributing to $\gamma_{b}(\bm H)$.  Further details of the fits are presented in the appendix.  Furthermore, our fitted value of the $m_{\downarrow}(0)\sim 180~m_e$ definitely characterizes URhGe as a heavy fermion compound.
It is important to note that the reentrant superconductivity in URhGe may arise from the increasing effective mass along the $b$-axis direction, which seems to be the most plausible scenario in light of the specific heat data in the context of our current model. Such field-dependent enhancements of $m_{b, \downarrow}$ should play a significant role in understanding the reentrant phase.

 In our single ellipsoidal FS model of the upper critical induction $B_{c2}(\theta,\phi,T)$,  $B_{c2}$ was embedded in a recursion relation that contained the three effective masses on that ellipsoidal FS\cite{Lorscher}.  In that model, we were only able to construct a theory of $B_{c2}$ for the low-field superconducting phase. Since our present work strongly suggests that $m_{b,\downarrow}({\bm H})$ is strongly peaked at $\mu_0H=12~$T, in order to construct a theory of $B_{c2}(\theta,\phi,T)$ for the reentrant phase, we need to use this double ellipsoidal FS model, with the Lorentzian field dependence of $m_{b,\downarrow}$ built into the embedded expressions, and solve self-consistently for the applied ${\bm H}$ and hence for $B_{c2}(\theta,\phi,T)$.  This will be a new type of calculation that has not previously been attempted, and we expect it to lead to new and very interesting behavior.

\begin{figure}
\center{\includegraphics[width=0.48\textwidth]{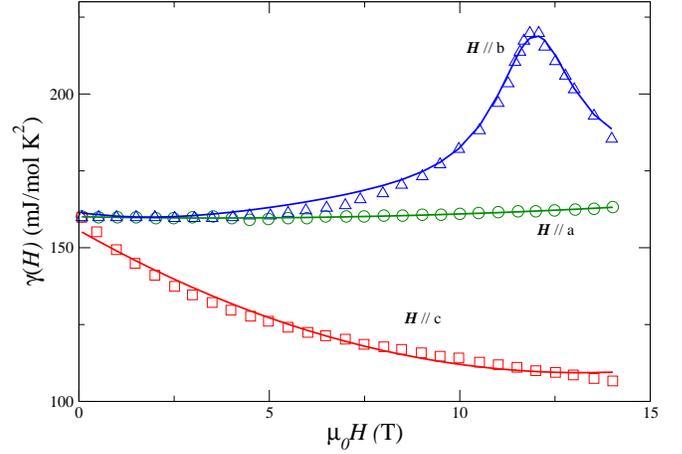}\vskip2pt
\caption{(color online) Experimental data for the linear $T$-coefficient of the specific heat along all three distinct crystal axes of URhGe, $\gamma_{a,b,c}$, provided by D. Aoki [16], and theoretical fits to the data for externally applied magnetic fields $\bm H$ along the $a$ (green), $b$ (blue), and $c$ (red) crystallographic directions. Solid color curves are fits of our model to the data.}}\label{fig1}
\end{figure}

\section{Summary and Conclusions}
We developed a model of the normal state of ferromagnetic superconductors consisting of two general ellipsoidal Fermi surfaces for the $\uparrow, \downarrow$ spin states that are ferromagnetically split in the presence of an arbitrarily oriented magnetic induction $\bm B$, which includes the spontaneous magnetization $\bm M_{0}$. By applying the Klemm-Clem transformations on each Fermi surface  separately, the problem was mapped onto one with two spherical Fermi surfaces with $\bm B_{\sigma}$ on each FS pointing along the crystal $z^{\prime}$-axis direction. We  calculated the linear $T$-coefficient of the specific heat, $\gamma(\bm B)$,  and obtained good fits to the experimental data of Aoki and Flouquet for URhGe. This model can be generalized to any number of ellipsoidal Fermi surfaces. Our results are expected to provide information crucial for the investigation of the high-field reentrant phase of the very strong candidate parallel-spin ferromagnetic superconductor URhGe.  In particular, this work strongly suggests that in order to construct a proper theory of $B_{c2}(\theta,\phi,T)$ for the reentrant phase of URhGe, and probably also for the $S$-shaped $B_{c2}(T)$ curve for ${\bm H}||\hat{\bm b}$ in UCoGe, one needs to solve self-consistently for ${\bm H}$ in terms of $m_{b,\downarrow}({\bm H})$, when ${\bm H}\cdot\hat{\bm b}\ne0$.

\section*{acknowledgments}
The authors are especially thankful to D. Aoki for kindly providing the specific heat data to fit, and also thank A. D. Huxley and K. Scharnberg for useful discussions.  This work was supported in part by the Florida Education Fund, the McKnight Doctoral Fellowship, the Specialized Research Fund for the Doctoral Program of Higher Education of China (no. 20100006110021) and by Grant no. 11274039 from the National Natural Science Foundation of China.

\section{Appendix}

We first calculate an expression for the field dependent chemical potential, $\mu_{\sigma}{(\bm B)}$, on the $\sigma$ Fermi surface, by calculating the particle density in the absence of a field, $n_{\sigma}(0)$, and in a field, $n_{\sigma}({\bm B})$. As in the Born-Sommerfeld approximation, in which the carrier density is assumed independent of $T$, we assume the total carrier density   $\sum_{\sigma} n_{\sigma}(0)=\sum_{\sigma} n_{\sigma}({\bm B})$ on both spin-split FSs to be independent of ${\bm B}$. In our formulation, we set $T\rightarrow0$.

We begin with the Hamiltonian, $\mathcal{H}$,  of our system
\begin{equation}\label{eq10}
\mathcal{H}=\sum_{j\sigma}\left[\frac{1}{2m_{j\sigma}({\bm B})}\left(k_{j}-eA_{j}\right)^{2}-\mu_{\sigma}(0)+g\mu_{B}{\bm \sigma}\cdot {\bm B}/2\right],
\end{equation}
where $\mu_{\sigma}(0)=\mu(0)-\sigma I/2$. We apply the combined anisotropic scale, rotation, and isotropic scale KC transformations
\begin{eqnarray}
x_{\mu}=\frac{1}{\alpha_{\sigma}\sqrt{\overline{m}_{\sigma,\mu}} }\sum_{\nu=1}^{3}\lambda_{\sigma,\nu\mu}\tilde{x}_{\sigma,\nu}, \\
\frac{\partial}{\partial x_{\mu}}=\alpha_{\sigma}\sqrt{\overline{m}_{\sigma,\mu}}\sum_{\nu=1}^{3}\lambda_{\sigma,\nu\mu}\frac{\partial}{\partial \tilde{x}_{\sigma,\nu}}, \\
B_{\mu}=\frac{1}{\alpha_{\sigma}\sqrt{\overline{m}_{\sigma,\mu}}} \sum_{\nu=1}^{3}\lambda_{\sigma,\nu\mu}\tilde{B}_{\sigma,\nu}, \\
H_{\mu}=\frac{1}{\alpha_{\sigma}\sqrt{\overline{m}_{\sigma,\mu}}} \sum_{\nu=1}^{3}\lambda_{\sigma,\nu\mu}\tilde{H}_{\sigma,\nu}, \\
A_{\mu}=\sqrt{\overline{m}_{\sigma,\mu}} \sum_{\nu=1}^{3}\lambda_{\sigma,\nu\mu}\tilde{A}_{\sigma,\nu},
\end{eqnarray}
where
\begin{eqnarray}
\mathbf{\lambda_{\sigma}}=\left(\begin{array}{ccc}
-\mathrm{sin}\phi^{\prime}_{\sigma} & -\mathrm{cos}\phi^{\prime}_{\sigma} & 0\\
\mathrm{cos}\theta^{\prime}_{\sigma}\mathrm{cos}\phi^{\prime}_{\sigma} & \mathrm{cos}\theta^{\prime}_{\sigma}\mathrm{sin}\phi^{\prime}_{\sigma} & -\mathrm{sin}\theta^{\prime}_{\sigma}\\
\mathrm{sin}\theta^{\prime}_{\sigma}\mathrm{cos}\phi^{\prime}_{\sigma} & \mathrm{sin}\theta^{\prime}_{\sigma}\mathrm{sin}\phi^{\prime}_{\sigma} & \mathrm{cos}\theta^{\prime}_{\sigma}
\end{array}\right),
\end{eqnarray}
and the transformed angles are given by
\begin{eqnarray}&&
\mathrm{cos}\theta^{\prime}_{\sigma}=\frac{\sqrt{\overline{m}_{3\sigma}}\mathrm{cos}\theta}{\alpha_{\sigma}\left(\theta,\phi\right)},  \\
&&\mathrm{sin}\theta^{\prime}_{\sigma}=\frac{\overline{\alpha}_{\sigma}\left(\phi\right)\mathrm{sin}\theta}{\alpha_{\sigma}\left(\theta,\phi\right)}, \\
&& \mathrm{cos}\phi^{\prime}_{\sigma}=\frac{\sqrt{\overline{m}_{1\sigma}}\mathrm{cos}\phi}{\overline{\alpha}_{\sigma}\left(\phi\right)}, \\
&& \mathrm{sin}\phi^{\prime}_{\sigma}=\frac{\sqrt{\overline{m}_{2\sigma}}\mathrm{sin}\phi}{\overline{\alpha}_{\sigma}\left(\phi\right)},
\end{eqnarray}
where $\overline{\alpha}_{\sigma}(\phi)=\alpha_{\sigma}(\pi/2,\phi)$.
We also obtain
\begin{eqnarray}
\epsilon_{\sigma}(k_{\sigma,\parallel},n+1/2)=\frac{k_{\sigma,\parallel}^{2}}{2m_{\sigma,\parallel}}+\omega_{\sigma,\perp}(n+1/2),
\end{eqnarray}
where $\omega_{\sigma, \perp}=\frac{eB}{m_{\sigma,\perp}({\bm B})}$, $m_{\sigma, \perp}({\bm B})=\frac{m_{\sigma}({\bm B})}{\alpha_{\sigma}\left(\theta,\phi\right)}$, and $m_{\sigma,\parallel}=m_{\sigma}\alpha_{\sigma}^{2}(\theta,\phi)$.

For the particle density at ${\bm B}=0$, we have
\begin{eqnarray}\label{eq11}
n_{\sigma}(0)=\int\frac{d^{3}k_{\sigma}^{\prime}}{\left(2\pi\right)^{3}}\Theta[\mu_{\sigma}(0)-\varepsilon_{\sigma}\left(k_{\sigma}^{\prime}
\right)],
\end{eqnarray}
where we have used the first KC scale transformation $k_{i}\rightarrow\sqrt{\overline{m}_{i\sigma}(0)}k_{i\sigma}^{\prime} $, which is different for each FS. We then obtain
\begin{equation}\label{eq12}
n_{\sigma}(0)= c_{\sigma}\int_{0}^{\mu_{\sigma}(0)}E_{\sigma}^{1/2}dE_{\sigma}=\frac{2m_{\sigma}^{3/2}(0)\mu_{\sigma}^{3/2}(0)}{3\pi^{2}\sqrt{2}},
\end{equation}
where $c_{\sigma}=\frac{m_{\sigma}^{3/2}(0)}{\pi^{2}\sqrt{2}}$, $\varepsilon_{\sigma}\left(k\right)=\sum_{j=1}^{3}\frac{k_{j\sigma}^{2}}{2m_{j\sigma}(0)}$,  $\varepsilon_{\sigma}\left(k_{\sigma}^{\prime}\right)=E_{\sigma}=\sum_{j=1}^{3}\frac{k_{j\sigma}^{\prime2}}{2m_{\sigma}(0)}$,  $m_{\sigma}(0)=[m_{1\sigma}(0)m_{2\sigma}(0)m_{3\sigma}(0)]^{1/3}$, and $m_{j\sigma}(0)$ are the single particle effective masses on each FS at ${\bm B}=0$. We sum over spin to obtain the total particle density for ${\bm B}=0$,
\begin{equation}\label{eq13}
\sum_{\sigma} n_{\sigma}(0)= \frac{2}{3\pi^{2}\sqrt{2}}\sum_{\sigma} \Pi_{\sigma}^{3/2}\left(0\right).
\end{equation}

Now, for ${\bm B}\ne0$,
\begin{eqnarray}\label{eq14}
&&n_{\sigma}(\bm B)=\frac{eB}{2\pi}\int\frac{dk_{\sigma,\parallel}}{2\pi}\nonumber\\
&&\times\sum_{n=0}^{\infty}\frac{1}{e^{\beta\left[k_{\sigma,\parallel}^{2}/2m_{\sigma, \parallel}({\bm B})+\omega_{\sigma, \perp}(n+1/2)-\mu_{\sigma}({\bm B})\right]}+1},
\end{eqnarray}
where we have applied the KC-transformations \cite{KC, book} on each ferromagnetically split FS to obtain $\omega_{\sigma, \perp}=\frac{eB}{m_{\sigma,\perp}({\bm B})}$, where $m_{\sigma,\perp}(\bm B)=m_{\sigma}({\bm B})/\alpha_{\sigma}(\theta,\phi)$ and $m_{\sigma, \parallel}({\bm B})=m_{\sigma}({\bm B})\alpha_{\sigma}^{2}(\theta,\phi)$.  To obtain the correct expression for  $m_{\sigma,||}({\bm B})$ from the KC transformations is non-trivial, but we checked this result by diagonalizing ${\cal H}$ with ${\bm A}$ chosen to lie in the $xy$, $yz$, and $xz$ planes. It also leads to the correct $B\rightarrow0$ limit of $m_{\sigma}(\bm B)=[m_{\sigma,\perp}^2({\bm B})m_{\sigma,||}(\bm B)]^{1/3}$.  We now obtain the expression for the particle density for ${\bm B}\ne0$,
\begin{eqnarray}\label{eq16}
n_{\sigma}(\bm B)&=&\frac{eB}{2\pi}\frac{\sqrt{m_{\sigma,\parallel}(\bm B)}}{\pi\sqrt{2}}\nonumber\\
&&\times\int E_{\sigma}^{-1/2}dE_{\sigma}\int_{0}^{\infty}dx\frac{1}{e^{\beta\left[E+\omega_{\sigma,\perp}x-\mu_{\sigma}(\bm B)\right]}+1}\nonumber\\
&&\times\left(1+2\sum_{s=0}^{\infty}\mathrm{cos}(2\pi sx)\right),
\end{eqnarray}
where we used the Poisson summation formula
\begin{equation}\label{eq17}
\sum_{n=0}^{\infty}f(n+1/2)=\int_{0}^{\infty}dxf(x)\left(1+2\sum_{s=0}^{\infty}(-1)^{s}\mathrm{cos}\left(2\pi sx\right)\right),
\end{equation}
and  made the change of variables $E_{\sigma}=\frac{k_{\sigma,\parallel}^{2}}{2m_{\sigma,\parallel}(\bm B)}$.

We calculate the non-oscillatory, $n_{\sigma}^{(1)}(\bm B)$, and oscillatory, $n_{\sigma}^{(2)}(\bm B)$, terms separately.
We first have
\begin{eqnarray}\label{eq18}
n_{\sigma}^{(1)}(\bm B)&=&\frac{eB}{2\pi}\frac{\sqrt{m_{\sigma,\parallel}(\bm B)}}{\pi\sqrt{2}}\frac{1}{\omega_{\sigma, \perp}}\int E_{\sigma}^{-1/2}dE_{\sigma}\nonumber\\
&&\times\int_{0}^{\infty}dy_{\sigma}\frac{1}{e^{\beta\left[E_{\sigma}+y_{\sigma}-\mu_{\sigma}(\bm B)\right]}+1},
\end{eqnarray}
which upon taking the $T\rightarrow0^{+}$  limit, becomes
\begin{eqnarray}\label{eq19}
n_{\sigma}^{(1)}(\bm B)&=&\frac{eB}{2\pi\omega_{\sigma, \perp}}\frac{\sqrt{m_{\sigma, \parallel}(\bm B)}}{\pi\sqrt{2}}\nonumber\\
&&\times\int_{0}^{\mu_{\sigma}(\bm B)}E_{\sigma}^{-1/2}dE_{\sigma}\int_{0}^{\mu_{\sigma}(\bm B)-E_{\sigma}}dy\nonumber\\
&&=\frac{\sqrt{2}m_{\sigma}^{3/2}(\bm B)\mu_{\sigma}^{3/2}(\bm B)}{3\pi^{2}},
\end{eqnarray}
where we changed the integration variable to $y_{\sigma}=\omega_{\sigma, \perp}x$, and we used the fact that $\sqrt{m_{\sigma,||}({\bm B})}m_{\sigma,\perp}({\bm B})=m_{\sigma}^{3/2}({\bm B})$. It is important to note that the exponential term in Equation (\ref{eq18}) becomes a theta-function when taking the zero-temperature limit.

Now we evaluate the oscillatory term, $n_{\sigma}^{(2)}(B)$, obtaining
\begin{eqnarray}\label{eq20}
n_{\sigma}^{(2)}(\bm B)&=&\frac{eB}{2\pi\omega_{\sigma, \perp}}\frac{\sqrt{m_{\sigma, \parallel}(\bm B)}}{\pi\sqrt{2}} \int_{0}^{\infty}dy_{\sigma}\nonumber\\
&&\left(\int_{0}^{\infty}E_{\sigma}^{-1/2}dE_{\sigma}\frac{1}{e^{\beta\left[E_{\sigma}+y_{\sigma}-\mu_{\sigma}(\bm B)\right]}+1}\right)\nonumber\\
&&\times2\sum_{s=1}^{\infty}(-1)^{s}\mathrm{cos}(2\pi sy_{\sigma}/\omega_{\sigma, \perp}).
\end{eqnarray}
We define
\begin{equation}\label{eq21}
f(y_{\sigma})=\int_{0}^{\infty}E_{\sigma}^{-1/2}dE_{\sigma}\frac{1}{e^{\beta\left[E_{\sigma}+y_{\sigma}-\mu_{\sigma}(\bm B)\right]}+1},
\end{equation}
which then leads to
\begin{eqnarray}\label{eq22}
n_{\sigma}^{(2)}(\bm B)&=&\frac{eB}{\pi\omega_{\sigma, \perp}}\frac{\sqrt{m_{\sigma, \parallel}(\bm B)}}{\pi\sqrt{2}}
\sum_{s=1}^{\infty}(-1)^{s}\nonumber\\
&&\times\int_{0}^{\infty}dy_{\sigma}f(y_{\sigma})\mathrm{cos}(2\pi sy_{\sigma})/\omega_{\sigma, \perp}).
\end{eqnarray}
We now evaluate the remaining integral by parts to arbitrary order and obtain
\begin{eqnarray}\label{eq23}
&&\int_{0}^{\infty}dy_{\sigma}f(y_{\sigma})\mathrm{cos}(2\pi sy_{\sigma}/\omega_{\sigma, \perp})=
\nonumber\\&&\sum_{n=0}^{\infty}(-1)^{n+1}\left(\frac{\omega_{\sigma, \perp}}{2\pi s}\right)^{2n+2}f^{(2n+1)}(0),
\end{eqnarray}
where $f^{(2n+1)}(0)=\frac{-(4n-1)!!}{2^{(2n)}\mu_{\sigma}^{(2n+1/2)}(\bm B)}$ is obtained from Eq. (\ref{eq21}) by letting $T\rightarrow0^{+}$, or $\beta\rightarrow+\infty$, so that the integration becomes
\begin{equation}\label{eq24}
f(y_{\sigma})=\int_{0}^{\mu_{\sigma}(\bm B)-y_{\sigma}}E_{\sigma}^{-1/2}dE_{\sigma}=2\sqrt{\mu_{\sigma}(\bm B)-y_{\sigma}},
\end{equation}
and by subsequently taking its $(2n+1)^{\rm th}$ derivative and evaluating the expressions at $y_{\sigma}=0$. Substituting Eq. (\ref{eq23}) into Eq. (\ref{eq22}), we obtain
\begin{eqnarray}\label{eq25}
n_{\sigma}^{(2)}(\bm B)&=&\frac{\sqrt{2}m_{\sigma}^{3/2}(\bm B)\mu_{\sigma}^{3/2}(\bm B)}{3\pi^{2}}\nonumber\\
&&\times\sum_{n=0}^{\infty}a_{n}\left(\frac{eB\alpha_{\sigma}(\theta,\phi)}{m_{\sigma}(\bm B)\mu_{\sigma}(\bm B)}\right)^{2n+2},
\end{eqnarray}
where $a_{n}=3(-1)^{n+1}(2-2^{-2n})\zeta(2n+2)\frac{(4n-1)!!}{(4\pi)^{2n+2}}$ and we
used $\sum_{s=1}^{\infty}(-1)^{s-1}\frac{1}{s^{2n+2}}=(1-2^{1-(2n+2)})\zeta(2n+2)$.

As in the Born-Sommerfeld approximation, we now equate the total particle density at ${\bm B}=0$, $\sum_{\sigma} n_\sigma(0)$, to the total particle density at ${\bm B}\ne0$, $\sum_{\sigma} [n_{\sigma}^{(1)}(\bm B)+n_{\sigma}^{(2)}(\bm B)]$, given by Eqs. (32) and (38), and obtain
\begin{eqnarray}\label{eq28}
&&\sum_{\sigma}\Pi_{\sigma}^{3/2}\left(0\right)=\nonumber\\
&&\sum_{\sigma}\Pi_{\sigma}^{3/2}\left(\bm B\right)\left(1+\sum_{n=0}^{\infty}a_{n}\left(\frac{eB\alpha_{\sigma}\left(\theta,\phi\right)}{\Pi_{\sigma}\left(\bm B\right)}\right)^{2n+2}\right).
\end{eqnarray}

We now present the details of the calculation of the linear $T$-coefficient of specific heat, $\gamma({\bm B})$, for an electron gas with a strong spin-split ellipsoidal Fermi surfaces, one for each spin projection  $\sigma=\{\uparrow,\downarrow\}$. Each ferromagnetically split Fermi surface has three distinct single particle effective masses, $\{m_{i\sigma}({\bm B})\}$.

Writing the entropy, $S$, of the system as in Eq.(\ref{eq5}) and taking the thermodynamic limit, $\sum_{\mathbf{k}\sigma}\rightarrow\sum_{\sigma}\frac{V}{\left(2\pi\right)^{3}}\int d^{3}\mathbf{k}$, and by applying the first KC transformation $k_{i\sigma}\rightarrow\sqrt{m_{i\sigma}/m_{\sigma}}k_{i\sigma}^{\prime}$, we obtain
\begin{eqnarray}\label{eq29}
S/V&=&k_{B}\sum_{\sigma}\int\frac{d^{3}\mathbf{k_{\sigma}^{\prime}}}{\left(2\pi\right)^{3}}\nonumber\\
&&\times\left[\frac{\beta\xi_{\sigma}\left(\mathbf{k_{\sigma}^{\prime}}\right)}{e^{\beta\xi_{\sigma}\left(\mathbf{k_{\sigma}^{\prime}}\right)}+1}+\mathrm{ln}\left(1+e^{-\beta\xi_{\sigma}\left(\mathbf{k_{\sigma}^{\prime}}\right)}\right)\right],
\end{eqnarray}
where $\xi_{\sigma}=\varepsilon\left(\mathbf{k_{\sigma}^{\prime}}\right)-\mu_{\sigma}$.

We then make use of the thermodynamic relationship for the specific heat $C_{V}=-\beta\frac{\partial\left(S/V\right)}{\partial\beta}$,  and obtain for $B=0,$
\begin{equation}\label{eq30}
C_{V}(0)=\frac{\beta^{2}k_{B}}{\sqrt{8}\pi^2}\sum_{\sigma}m_{\sigma}^{3/2}(0)\int_{-\mu_{\sigma}(0)}^{\infty}\frac{\xi_{\sigma}^{2}d\xi_{\sigma}\sqrt{\xi_{\sigma}+\mu_{\sigma}(0)}}{1+\mathrm{cosh}(\beta\xi_{\sigma})}.
\end{equation}
Upon introducing the change of variables $x_{\sigma}=\beta\xi_{\sigma}$, allowing $T\rightarrow0^{+}$, and making use of the common integral $\int_{-\infty}^{\infty}\frac{x^{2}dx}{1+\mathrm{cosh}x}=\frac{2\pi^{2}}{3}$, we find the $B=0$ $T$-coefficient to be $\gamma(0)=\sum_{\sigma}\frac{m_{\sigma}^{3/2}(0)\mu_{\sigma}^{1/2}(0)}{3\sqrt{2}}$, the standard textbook result.

Now we generalize our result to  finite ${\bm B}$, using
\begin{eqnarray}\label{eq1_1}
C_{V}({\bm B})&=&\frac{\beta^{2}k_{B}}{2}\frac{eB}{2\pi}\sum_{\sigma}\int\frac{dk_{\sigma,\parallel}}{\left(2\pi\right)}\nonumber\\
&&\times\sum_{n=0}^{\infty}\frac{\xi_{\sigma}^{2}\left(k_{\sigma,\parallel},n+1/2\right)}{1+\mathrm{cosh}\left(\beta\xi_{\sigma}\left(k_{\sigma,\parallel},n+1/2\right)\right)},
\end{eqnarray}
where
\begin{equation}\label{eq31}
\xi_{\sigma}\left(k_{\sigma,\parallel},n+1/2\right)=\frac{k_{\sigma,\parallel}^{2}}{2m_{\sigma,\parallel}(\bm B)}+\omega_{\sigma, \perp}(n+1/2)-\mu_{\sigma}(\bm B).
\end{equation}
 We use the Poisson summation formula, and obtain  non-oscillatory and oscillatory terms, $C_{V}^{(1)}(\bm B)$ and $C_{V}^{(2)}(\bm B)$, respectively, which we  evaluate separately.  We find
\begin{eqnarray}\label{eq32}
C_{V}^{(1)}(\bm B)&=&\frac{\beta^{2}k_{B}}{\sqrt{8}\pi}\frac{eB}{2\pi}\sum_{\sigma}\frac{m_{\sigma,||}^{1/2}({\bm B})}{\omega_{\sigma, \perp}}\int_{0}^{\infty}E_{\sigma}^{-1/2}dE_{\sigma}\nonumber\\
&&\times\int_{0}^{\infty}dy_{\sigma}\left[\frac{\left(E_{\sigma}-\mu_{\sigma}(\bm B)+y_{\sigma}\right)^{2}}{1+\mathrm{cosh}\beta\left(E_{\sigma}-\mu_{\sigma}(\bm B)+y_{\sigma}\right)}\right],\nonumber\\
\end{eqnarray}
where we used the change of variables $y_{\sigma}=\omega_{\sigma, \perp}x$.
We now let $\xi_{\sigma}=E_{\sigma}-\mu_{\sigma}({\bm B})+y_{\sigma}$, and $\beta\xi_{\sigma}=x_{\sigma}$, and obtain
\begin{eqnarray}\label{eq33}
C_{V}^{(1)}(\bm B)&=&\frac{k_{B}^{2}T}{\sqrt{8}\pi}\frac{eB}{2\pi}\sum_{\sigma}\frac{m_{\sigma,||}^{1/2}({\bm B})}{\omega_{\sigma, \perp}}\int_{0}^{\infty}c_{1\sigma}E_{\sigma}^{-1/2}dE_{\sigma}\nonumber\\
&&\times\int_{\beta[-\mu_{\sigma}(\bm B)+E_{\sigma}]}^{\infty}dx_{\sigma}\left[\frac{x_{\sigma}^{2}}{1+\mathrm{cosh}x_{\sigma}}\right].
\end{eqnarray}
We then take the zero temperature limit, $T\rightarrow0^{+}$, for which $\beta[-\mu_{\sigma}(\bm B)+E_{\sigma}]\rightarrow -\infty$,  and evaluating the energy integral up to the chemical potential in a field, $\int_{0}^{\mu_{\sigma}(\bm B)}E_{\sigma}^{-1/2}dE_{\sigma}$, we obtain the non-oscillatory part,
\begin{equation}\label{eq34}
C_{V}^{(1)}(\bm B)=\frac{k^{2}_{B}T}{3\sqrt{2}}\sum_{\sigma}m_{\sigma}^{3/2}(\bm B)\mu_{\sigma}^{1/2}(\bm B).
\end{equation}

Now we evaluate the oscillatory second term, obtaining
\begin{eqnarray}\label{eq35}
C_{V}^{(2)}(\bm B)&=&\frac{\beta^{2}k_{B}eB}{\sqrt{8}\pi^2}\sum_{\sigma}\frac{m_{\sigma,||}^{1/2}({\bm B})}{\omega_{\sigma,\perp}}\sum_{s=1}^{\infty}(-1)^{s}\nonumber\\
&&\times\int_{0}^{\infty}dy_{\sigma}f(y_{\sigma})\mathrm{cos}\left(2\pi s\frac{y_{\sigma}}{\omega_{\sigma,\perp}}\right),
\end{eqnarray}
where
\begin{equation}\label{eq36}
f(y_{\sigma})\equiv\int_{0}^{\infty}E_{\sigma}^{-1/2}dE_{\sigma}\frac{[E_{\sigma}-\mu_{\sigma}(\bm B)+y_{\sigma}]^{2}}{1+\mathrm{cosh}\beta\left(E_{\sigma}-\mu_{\sigma}(\bm B)+y_{\sigma}\right)}.
\end{equation}
By successively integrating by parts to infinite order, we obtain the infinite series expansion,
\begin{eqnarray}\label{eq38}
&&\int_{0}^{\infty}f(y_{\sigma})\mathrm{cos}\left(2\pi sy_{\sigma}/\omega_{\sigma,\perp}\right)dy_{\sigma}=\nonumber\\
&&\sum_{n=0}^{\infty}(-1)^{n+1}\frac{\omega_{\sigma,\perp}^{2n+2}}{\left(2\pi s\right)^{2n+2}}f^{(2n+1)}(0),\\
&&f^{(2n+1)}(0)=\frac{2\pi^{2}}{3\beta^{3}}\frac{1}{2^{2n+1}}\left(4n+1\right)!!\frac{1}{[\mu_{\sigma}(\bm B)]^{2n+3/2}},
 \end{eqnarray}
 which we obtained from Eq. (\ref{eq36}) by integrating with respect to $x_{\sigma}=\beta[E_{\sigma}-\mu_{\sigma}(B)-y_{\sigma}]$, and then allowing $T\rightarrow0^{+}$,
or equivalently $\beta\rightarrow+\infty$.  This results in
\begin{equation}\label{eq39}
f(y_{\sigma})=\frac{2\pi^{2}}{3\beta^{3}}[\mu_{\sigma}(\bm B)-y_{\sigma}]^{-1/2},
\end{equation}
which can then be differentiated $(2n+1)$-times to obtain the desired result. Substituting Eq. (\ref{eq39}) into Eq. (\ref{eq35}), we obtain
\begin{eqnarray}\label{eq40}
C_{V}^{(2)}(\bm B)&=&\frac{k_{B}eB}{\sqrt{2}\pi^2}\left(\frac{\pi^{2}}{3\beta}\right)\sum_{\sigma}\frac{m_{\sigma,||}^{1/2}({\bm B})}{\omega_{\sigma,\perp}}\sum_{n=0}^{\infty}(-1)^{n}\nonumber\\
&&\times\sum_{s=1}^{\infty}\frac{(-1)^{s-1}}{s^{2n+2}}\frac{\omega_{\sigma,\perp}^{2n+2}}{(2\pi)^{2n+2}}\frac{(4n+1)!!}{2^{2n+1}}\nonumber\\
&&\times\frac{1}{\left[\mu_{\sigma}(\bm B)\right]^{2n+3/2}},
\end{eqnarray}
so that we have
\begin{eqnarray}\label{eq41}
C_{V}^{(2)}(\bm B)&=&\frac{k_B^2T}{3\sqrt{2}}\sum_{\sigma}m_{\sigma}^{3/2}(\bm B)\mu^{1/2}_{\sigma}(\bm B)\nonumber\\
&&\times\sum_{n=0}^{\infty}\lambda_{n}\left(\frac{eB\alpha_{\sigma}(\theta,\phi)}{m_{\sigma}(\bm B)\mu_{\sigma}(\bm B)}\right)^{2n+2},
\end{eqnarray}
where
\begin{eqnarray}\label{eq42}
\lambda_{n}&=&\frac{1}{3}\left(4n+1\right)a_{n}.
\end{eqnarray}
\vskip5pt

Combining $C_{V}(\bm B)=C_{V}^{(1)}(\bm B)+C_{V}^{(2)}(\bm B)$, we obtain
\begin{eqnarray}\label{eq43}
\gamma\left({\bm B}\right)&=&\frac{k^{2}_{B}}{3\sqrt{2}}\sum_{\sigma}m_{\sigma}\left(\bm B\right)\Pi_{\sigma}^{1/2}\left(\bm B\right)\nonumber\\
&&\times\left(1+\sum_{n=0}^{\infty}\lambda_{n}\left(\frac{eB\alpha_{\sigma}\left(\theta,\phi\right)}{\Pi_{\sigma}\left(\bm B\right)}\right)^{2n+2}\right).
\end{eqnarray}

Here we present the derivation for the conversion factor $\lambda$ from m\textsuperscript{3} to moles, the effective mass in zero field $m_{\downarrow}(0)$, and the chemical potential in zero field $\mu_{\downarrow}(0)$.

The number $N_{\rm U}$ of Uranium atoms per unit cell is 4 \cite{Aoki_Flouquet}, the volume of a unit cell is $V_{\rm c}=4.35\times 10^{-10}\times6.90\times10^{-10}\times7.52\times10^{-10}\textnormal{m\textsuperscript{3}}=2.25\times10^{-28}\textnormal{m\textsuperscript{3}}/\textnormal{unit cell}$ \cite{Tran}, and there are Avogadro's  number of atoms in one mole, $N_A=6.022\times10^{23}\textnormal{U atoms/mol of URhGe}$. Combining these numbers,  we  obtain the conversion factor
\begin{equation}
\lambda=(V_{\rm c}/N_{\rm U})\times N_{A}=3.4 \times 10^{-5} \textnormal{m\textsuperscript{3}/mol}.
\end{equation}
We also have from Yelland $et$ $al.$ \cite{Yelland} that the number of carriers per Uranium is
\begin{equation}
N_{\textnormal{c}}/\textnormal{U}=2.1\times10^{-3}\textnormal{carriers/U},
\end{equation}
from which we calculate $k_F$, using
\begin{equation}
N_{\textnormal{c}}/\textnormal{U}=2.1\times 10^{-3}=\frac{V_{c}\times k^{3}_{F}}{6\pi^{2}},
\end{equation}
and obtain
\begin{equation}
k_{F}=0.82\times 10^{10} \textnormal{/m}.
\end{equation}

Now, we substitute our values into the expression
\begin{equation}
\gamma(0)=\frac{2(\alpha m_{e})k_{F}k^{2}_{B}\times \lambda}{\hbar^{2}},
\end{equation}
and obtain a dimensionless number $\alpha$, where $m_{\downarrow}(0)=\alpha m_e$, which we would expect to be large for heavy Fermion materials such as URhGe.  In our fit, we found $m_{\downarrow}(0)=180~m_{e}$, or $\alpha\approx 180$,
where we have used $\gamma(0)=0.16~\textnormal{J/mol K$^2$}$, and we  used $\lambda$ to convert from moles to m\textsuperscript{3}, and the factor of $2$ comes from taking both spin contributions.

Now we calculate the chemical potential on the $\downarrow$ FS in zero field
\begin{equation}
\gamma(0)=\frac{k^2_{B}}{3\hbar^{3}\sqrt{2}}\lambda m^{3/2}_{\downarrow}(0)\mu^{1/2}_{\downarrow}(0),
\end{equation}
and obtain
\begin{equation}
\mu_{\downarrow}(0)=3.2\times 10^{-21}\textnormal{J},
\end{equation}
which is 20 meV, appropriate for a heavy fermion compound with $m_{\downarrow}(0)=180~m_e$.

We also obtained good least-squares fits to the data for ${\bm H} \parallel \hat{\bm a}$ and ${\bm H} \parallel \hat{\bm c}$ shown in Fig. 3.  We found $\gamma_{\bm H \parallel a}=160~\textup{mJ/(mol K\textsuperscript{2})}-7.0~\textup{mJ/(mol K\textsuperscript{2} T)}(\mu_{0}H)+0.27~\textup{mJ/(mol K\textsuperscript{2}T\textsuperscript{2})}(\mu_{0}H)^{2}$, and $\gamma_{\bm H \parallel c}=160~\textup{mJ/(mol K\textsuperscript{2})}-0.85~\textup{mJ/(mol K\textsuperscript{2} T)}(\mu_{0}H)+0.37~\textup{mJ/(mol K\textsuperscript{2} T\textsuperscript{2})}(\mu_{0}H)^2$.  These functions are respectively shown as the green and red curves in Fig. 3.

\end{document}